\documentclass[usenatbib]{mn2e}
\usepackage{amsmath, amssymb}
\usepackage{amsfonts}
\usepackage{epsfig,rotating}

\def\simeq{
\mathrel{\raise.3ex\hbox{$\sim$}\mkern-14mu\lower0.4ex\hbox{$-$}}
}

\def\ltsima{$\; \buildrel < \over \sim \;$}
\def\simlt{\lower.5ex\hbox{\ltsima}}
\def\gtsima{$\; \buildrel > \over \sim \;$}
\def\simgt{\lower.5ex\hbox{\gtsima}}

\def\msun{{\rm M_{\odot}}}

\def\be{\begin{equation}}
\def\ee{\end{equation}}

\def\del#1{{}}
\def\ltsima{$\; \buildrel < \over \sim \;$}
\def\simlt{\lower.5ex\hbox{\ltsima}}
\def\gtsima{$\; \buildrel > \over \sim \;$}
\def\simgt{\lower.5ex\hbox{\gtsima}}

\newcommand{\apj}{ApJ}

\newcommand{\mnras}{MNRAS}
\newcommand{\aap}{A\&A}
\newcommand{\araa}{ARA\&A}
\newcommand{\apjl}{ApJL}

\newcommand{\nat}{Nature}

\newcommand{\pasj}{PASJ}

\title[Massive outflows suggest AGN fade slowly]{Massive outflow properties suggest AGN fade slowly}
\author[Kastytis Zubovas]{Kastytis Zubovas$^{1,2,\star}$ \\
  $^{1}$Center for Physical Sciences and Technology, Saul\.{e}tekio av. 3, Vilnius LT-10257, Lithuania \\
  $^{2}$Vilnius University Observatory, Saul\.{e}tekio av. 9, Bldg III, Vilnius LT-10222, Lithuania\\
  $^{\star}$ {E-mail:~} {\rm kastytis.zubovas@ftmc.lt} }

\begin{document}

\maketitle

\begin{abstract}
  Massive large-scale AGN outflows are an important element of galaxy
  evolution, being a way through which the AGN can affect most of the
  host galaxy. However, outflows evolve on timescales much longer than
  typical AGN episode durations, therefore most AGN outflows are not
  observed simultaneously with the AGN episode that inflated them. It
  is therefore remarkable that rather tight correlations between
  outflow properties and AGN luminosity exist. In this paper, I show
  that such correlations can be preserved during the fading phase of
  the AGN episode, provided that the AGN luminosity evolves as a power
  law with exponent $\alpha_{\rm d} \sim 1$ at late times. I also show
  that subsequent AGN episodes that illuminate an ongoing outflow are
  unlikely to produce outflow momentum or energy rates rising above
  the observed correlations. However, there may be many
  difficult-to-detect outflows with momentum and energy rates lower
  than expected from the current AGN luminosity. Detailed observations
  of AGN outflow properties might help constrain the activity
  histories of typical and/or individual AGN.
\end{abstract}


\section{Introduction}

Kiloparsec-scale outflows of molecular gas have been observed in
numerous galaxies \citep{Feruglio2010A&A, Rupke2011ApJ, Sturm2011ApJ,
  Alatalo2011ApJ, Cicone2014A&A}, with mass outflow rates $\dot{M} >
1000 \;\msun $~yr$^{-1}$ and velocities $v_{\rm out} >
1000$~km~s$^{-1}$. The properties of these outflows are well explained
by the AGN wind-driven outflow model \citep{King2003ApJ,
  King2010MNRASa, King2015ARA&A}, which posits that a mildly
relativistic wind originating in the AGN accretion disc inflates a
large-scale outflow by transferring its kinetic energy rate
$\dot{E}_{\rm kin} \simeq 0.05 L_{\rm AGN}$ to the surrounding
interstellar medium (ISM). The velocities, kinetic energy and momentum
rates of the observed outflows \citep{Cicone2014A&A} agree very well
with analytical predictions \citep{Zubovas2012ApJ}.  In at least one
case \citep{Tombesi2015Natur}, both a small-scale relativistic wind
and a large-scale molecular outflow are detected in the same galaxy,
with their kinetic energies agreeing with analycial predictions
\citep[however, see ][for a re-examination of these results, finding a
  smaller average outflow mass, momentum and energy rate over a longer
  flow timescale of $\sim 10$~Myr]{Veilleux2017ApJ}.

These close correlations are typically considered evidence that winds
of the observed AGN drive the observed outflows. However, on closer
inspection, this simple interpretation is unlikely to be correct. AGN
winds are highly intermittent, with lifetimes of only a few months
\citep{King2015ARA&A}. AGN are observed to change their luminosity
significantly on timescales from from several years
\citep{Gezari2017ApJ} to several times $10^4$~yr
\citep{Keel2017ApJ}. The crossing timescale of the outflow is,
however, $t_{\rm dyn, out} \sim R_{\rm out}/v_{\rm out} \sim 1 R_{\rm
  kpc} v_3^{-1}$~Myr, with $R_{\rm out} \equiv 1 R_{\rm kpc}$~kpc and
$v_{\rm out} \equiv 10^3 v_3$~km~s$^{-1}$. Therefore, the outflows
that are observed at distances beyond $\sim 0.1$~kpc from the galactic
nucleus are unlikely to have been inflated by the current AGN
episode. In fact, it is expected that an outflow would persist for an
order of magnitude longer than the duration of the AGN phase that
inflated it \citep{King2011MNRAS}. During this time, other AGN
episodes might occur in the galaxy, but there is no {\em a priori}
reason why these should have luminosity values correlating as well
with the outflow properties as observed ones do.

In this paper, I investigate the possible solutions to this issue,
considering outflow properties as potential constraints on the AGN
luminosity evolution. First of all, in Section \ref{sec:model}, I
briefly review the AGN wind outflow model and present the
possibilities for maintaining correlations between outflows and
AGN. The most promising possibility is that on long timescales, the
AGN luminosity tends to vary in such a way that correlations are
preserved. I investigate this possibility in Section \ref{sec:sims},
finding that if the AGN luminosity fades as a power law over time, as
suggested by some evolutionary models, the correlations with outflow
properties can be preserved during the fading phase as well as the
driving phase. I discuss the implications of the results in Section
\ref{sec:discuss} and conclude in Section \ref{sec:concl}.

\section{Outflows in fading AGN} \label{sec:model}

\subsection{Wind outflow model}

The model of AGN outflows driven by accretion disc winds was proposed
by \citet{King2003ApJ} in order to explain the observed $M-\sigma$
correlation \citep{Ferrarese2000ApJ}. It is based on the fact that
accretion disc radiation can drive a quasi-relativistic ($v_{\rm w}
\sim 0.1c$) wide-angle wind, also known as an ultra-fast outflow
(UFOs). The wind tends to self-regulate to keep the outflow rate
$\dot{M}_{\rm w}$ similar to the SMBH accretion rate
\citep{King2010MNRASa}. The wind energy rate is then
\begin{equation} \label{eq:winden}
  L_{\rm w} = \frac{\dot{M}_{\rm w} v_{\rm w}^2}{2} =
  \frac{\eta}{2}L_{\rm AGN} \simeq 0.05 L_{\rm AGN},
\end{equation}
where $\eta \simeq 0.1$ is the radiative efficiency of accretion.

The wind shocks against the surrounding interstellar medium (ISM) and
drives a large-scale outflow. Depending on the distance between the
AGN and the wind shock, the wind might be cooled efficiently by the
AGN radiation field via the inverse-Compton process, leaving only the
wind momentum to push against the gas. However, this critical `cooling
radius' is probably so small as to be irrelevant
\citep{Faucher2012MNRASb, Bourne2013MNRAS}, and the shocked wind is
always essentially adiabatic. In this case, the outflow is
energy-driven - it has the same energy rate as the wind
(eq. \ref{eq:winden}). Such an outflow is capable of removing most of
the gas from the galaxy spheroid. The predicted relation between the
properties of the outflow and the AGN luminosity agree very well with
observations \citep{Zubovas2012ApJ, Cicone2014A&A}. For a more
detailed review of the model, see \citet{King2015ARA&A}.

The model, as presented above, assumes that the AGN luminosity is
fixed at either some constant fraction of $L_{\rm Edd}$ or at
zero. However, this is clearly unrealistic: if an outside reservoir
ceases feeding the central SMBH, its accretion disc gets depleted on a
viscous timescale, and the AGN luminosity should decrease over a
finite amount of time as well, perhaps of order $10^4-10^5$~yr
\citep{Schawinski2010ApJb, Ichikawa2016PASJ, Keel2017ApJ}. The
outflow, meanwhile, continues to expand and coast
\citep{King2011MNRAS}. Before it dissipates completely, several more
AGN episodes can occur, with potentially different maximum
luminosities. The existence of AGN-outflow correlations suggests
either that outflows can only be observed during the driving AGN
phase, or that correlations between AGN luminosity and outflow
properties are maintained during the fading phase and during
subsequent AGN episodes.

\subsection{Outflow observability} \label{sec:observ}

The simplest solution to the problem would be that outflows are only
observed at times when illuminated by powerful AGN, and become
unobservable almost as soon as the AGN begins fading. That way, all
observed outflows would be inflated by the AGN of the same luminosity
as observed, and thus their properties would correlate with the AGN
luminosity. This may be the case for UFOs. Their sizes are $\ll 1$~pc
\citep{Tombesi2012MNRAS, Tombesi2013MNRAS}, giving a light-travel time
of $< 1$~yr. Therefore, one might expect UFOs to dissipate, recombine
and become essentially unobservable in tandem with the fading of the
central source, on timescales as short as several tens of years
\citep{Ichikawa2017arXiv}.

Massive kpc-scale outflows, on the other hand, should persist and be
detectable long after the AGN fades. The shock-ionised gas in the
outflows recombines and cools down rapidly \citep{Zubovas2014MNRASa},
leading to somewhat lower typical momentum and kinetic energy rates of
ionised outflows compared with molecular ones \citep{Fiore2017A&A}.
However, even ionised gas can be detectable on timescales of several
times $10^4$~yr after the AGN fades \citep{Keel2017ApJ}. Outflowing
molecular gas should remain kinematically distinct from the rest of
the galactic material for a long time, up to an order of magnitude
longer than the duration of the driving AGN phase
\citep{King2011MNRAS}, assuming that it was kinematically distinct in
the first place, i.e. that the outflow velocity was much higher than
the gas velocity dispersion in the host galaxy.

At some point after the AGN begins fading, it becomes unobservable.
Then the remaining outflow might be identified as being driven by star
formation, since no AGN would be visible in the host galaxy of the
outflow. However, during the intermediate fading phase, the outflow
can still be identified as being driven by the AGN. It seems unlikely,
given the evidence above, that the AGN outflow would become
undetectable very soon after the AGN begins dimming.

\subsection{Conditions for outflow existence} \label{sec:coinc}

Another reason limiting the possible variations of the correlations is
a sort of cosmic coincidence. The presence of massive outflows
requires two conditions that limit the time range during which they
can appear.

First of all, the driving AGN must be luminous enough. The critical
luminosity is $L_{\rm crit} \simeq L_{\rm Edd}\left(M_\sigma\right)$,
where $M_\sigma \simeq 3.68 \times 10^8 \sigma_{200}^4 \msun$ is the
critical SMBH mass, corresponding rather well to the observed
$M-\sigma$ relation \citep{King2010MNRASa, Zubovas2012MNRASb}. This
means that massive outflows can only occur once the SMBH grows beyond
a critical mass \citep{King2005ApJ}, i.e. the presence of the outflow
implies that the SMBH is already on the $M-\sigma$ relation or close
to it.

On the other hand, a molecular outflow can only form if the host
galaxy contains large amounts of gas. However, the outflow sweeps that
gas out of the host, so any subsequent significant episodes of AGN
activity are unlikely to produce massive outflows. In fact, our Galaxy
may be evidence of this: the Fermi bubbles \citep{Su2010ApJ} are most
likely inflated by a recent AGN episode \citep{Zubovas2012MNRASa} and
are thus analogous to the outflows in other galaxies. However, due to
the low gas mass in the bulge and halo of the Milky Way, the bubbles
are only visible in gamma rays and would be almost impossible to
detect in other galaxies, even though they should exist in the
majority of them \citep{King2011MNRAS}. In fact, it was only very
recently that Fermi bubble analogues were detected in M31
\citep{Pshirkov2016MNRAS}.

These two constraints strongly suggest that we only observe outflows
during a very particular time in a given galaxy's evolution - the time
when the central SMBH has recently reached its mass as given by the
$M-\sigma$ relation and is now driving an outflow that would shut off
both star formation in the spheroid and subsequent SMBH growth.

Even if this is the case, the outflow persists an order of magnitude
longer than the AGN episode which inflated it \citep{King2011MNRAS}.
During this time, there may be several more AGN episodes in the host
galaxy, which can illuminate the existing outflow and be observed
simultaneously. Even during a single AGN episode, the correlation with
outflow properties can in principle be broken. A typical AGN phase
probably lasts only $t_{\rm q} \sim 5 \times 10^4 - 2\times 10^5$~yr
\citep{Schawinski2015MNRAS, King2015MNRAS}, followed by a fading phase
of duration $t_{\rm d} \sim$ a few times$ 10^4$~yr
\citep{Keel2017ApJ}. During this phase, the correlations between
outflow and AGN properties might be disrupted. However, the time for
which the outflow properties would appear abnormally large compared
with the AGN luminosity is $> t_{\rm d}$, giving a probability of
observing abnormally large outflow properties $p > t_{\rm
  d}/\left(t_{\rm d} + t_{\rm q}\right) \sim 0.5$. Given that
abnormally large values are not commonly observed, either $t_{\rm
  d}$ must be very short, so that AGN fade away extremely abruptly,
or AGN luminosity must decrease slower than exponentially. The first
possibility appears unlikely, since the AGN is fed via a disc, which
takes a finite time to dissipate.

\subsection{Viscous evolution of AGN feeding rate} \label{sec:visc}

AGN accretion discs evolve and are depleted on a viscous timescale
\begin{equation}
  t_{\rm visc} \sim \frac{1}{\alpha_{\rm
      visc}}\left(\frac{H}{R}\right)^{-2} t_{\rm dyn} \sim 600 M_8
  \left(\frac{R}{R_{\rm ISCO}}\right)^{3/2} {\rm yr}.
\end{equation}
Here, $M_8 \equiv M_{\rm SMBH} / 10^8 \msun$ and $R_{\rm ISCO} \sim 3
R_{\rm Schw}$ is the radius of the innermost stable circular
orbit. For a $10^8 \msun$ black hole, the viscous timescale at the
outer edge of the accretion disc, which is at $R \simeq 0.01$~pc$
\simeq 700 R_{\rm ISCO}$ \citep{King2007MNRAS}, is
\begin{equation}
  t_{\rm visc, 0.01 {\rm pc}} \simeq 1.2 \times 10^8 {\rm yr}.
\end{equation}
The characteristic disc evolution timescale should lie somewhere
between these two extremes. \citet{King2015MNRAS} estimate that the
typical duration of an AGN event should be a few times $10^5$~yr. One
can then assume that the accretion disc is depleted and the AGN
luminosity fades on a similar timescale $t_{\rm q}$. The precise shape
of the luminosity function is then \citep{King2007MNRAS}
\begin{equation} \label{eq:ext_powerlaw}
  L_{\rm AGN}\left(t\right) = L_{\rm Edd}\left(1+\frac{t}{t_{\rm
      q}}\right)^{-19/16},
\end{equation}
which shows a power-law decay at late times.

A power-law decay of AGN luminosity seems a promising possibility for
keeping the outflow properties correlating with $L_{\rm AGN}$, since
analytical calculations show that stalling outflow properties also
evolve as power laws in time \citep{King2011MNRAS}.

\subsection{Section summary}

The arguments given above suggest that:
\begin{itemize}
\item AGN outflows should remain detectable while the AGN fades and
  long afterward, although they will not necessarily be identified as
  AGN-driven outflows if the AGN is too weak;
\item There is a non-negligible probability of detecting an outflow in
  a galaxy with a fading AGN;
\item The AGN luminosity is likely to decay as a power law at late
  times.
\end{itemize}

In order to determine whether the outflow correlations can be
maintained in such systems, I performed a numerical investigation of
outflow propagation in galaxies with flickering AGN. The results are
presented in the next Section.

\section{Simulated outflows in fading AGN} \label{sec:sims}

\subsection{Numerical model} \label{sec:nummodel}

\begin{figure*}
  \centering
    \includegraphics[trim = 0 0 0 0, clip, width=\textwidth]{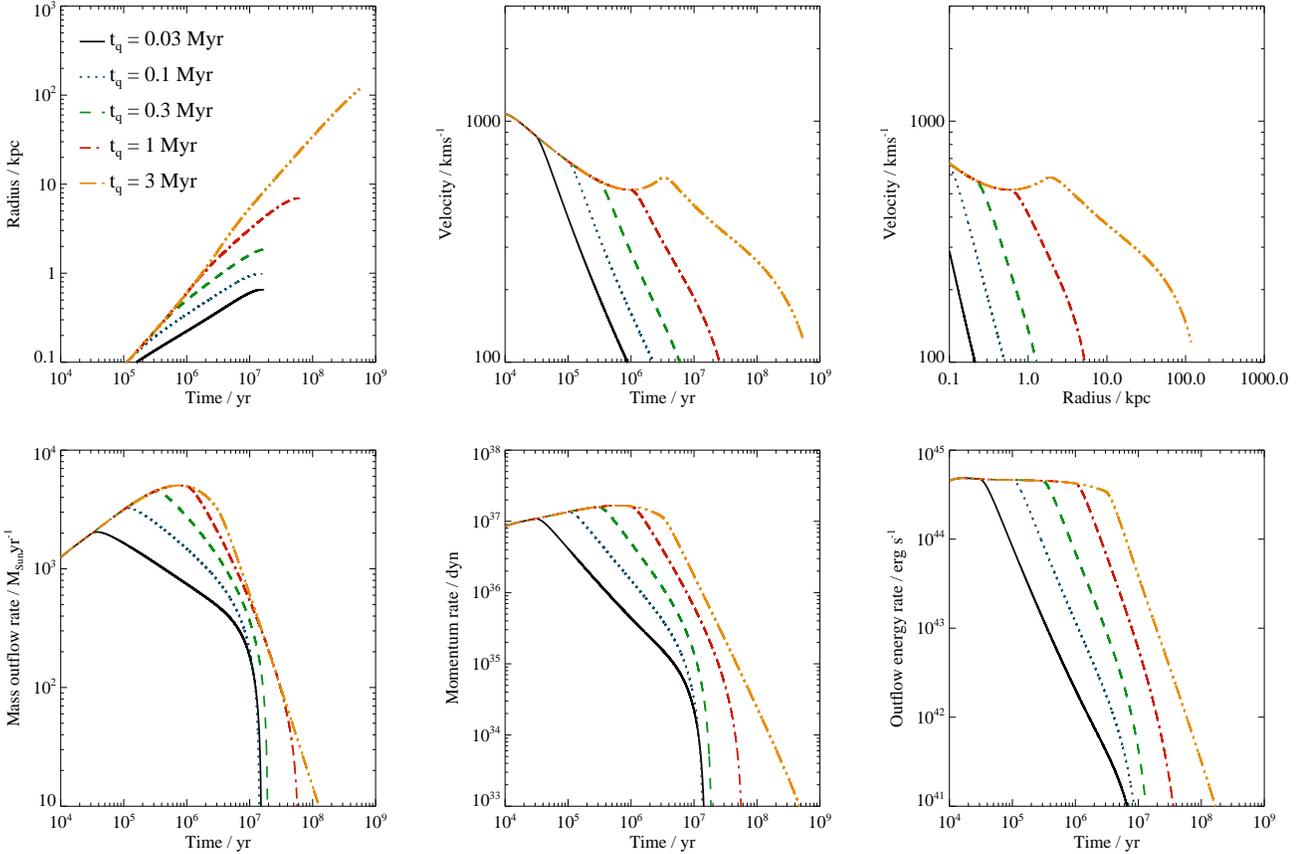}
  \caption{Evolution of outflow parameters in simulations with a
    single activity episode, followed by immediate complete shutdown
    of the AGN. From left to right, top to bottom, the panels show
    time evolution of radius, time evolution of velocity, velocity
    against radius, time evolution of the mass outflow rate, time
    evolution of the momentum rate and time evolution of the outflow
    kinetic energy rate. The five simulations have different quasar
    activity durations, as labelled in the top-left panel.}
  \label{fig:constdrop}
\end{figure*}

\begin{figure*}
  \centering
    \includegraphics[trim = 0 0 0 0, clip, width=\textwidth]{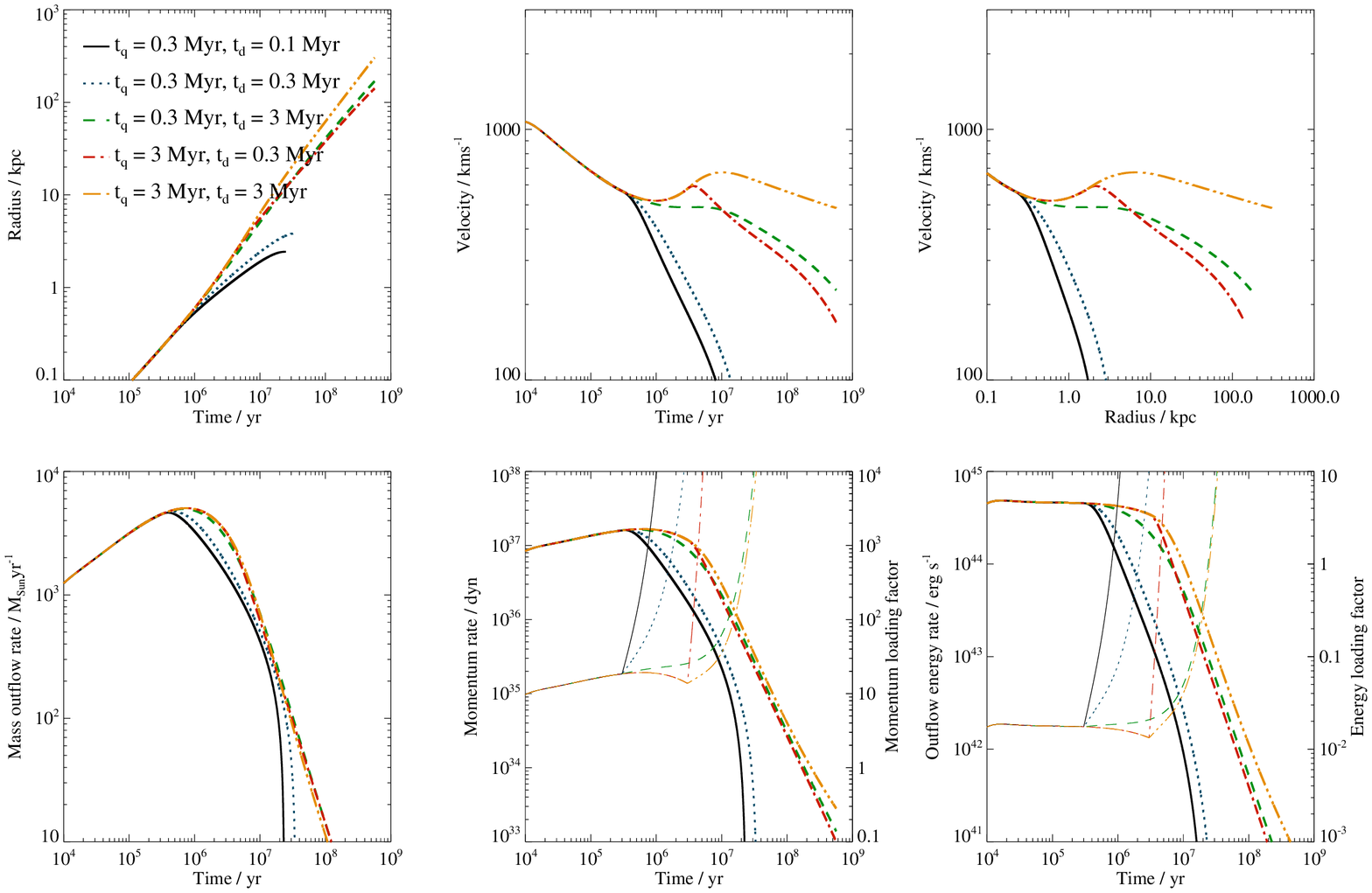}
  \caption{Same as Figure \ref{fig:constdrop}, but for simulations
    with exponential decay of AGN luminosity. Thin lines in the last
    two panels show the ratio between outflow momentum (bottom middle)
    or kinetic energy (bottom right) rate and the corresponding AGN
    parameter, $L_{\rm AGN}/c$ or $L_{\rm AGN}$ respectively, with
    scale on the right. Values of the free parameters $t_{\rm q}$ and
    $t_{\rm d}$ are labelled in the top-left panel.}
  \label{fig:expdrop}
\end{figure*}

I numerically integrate the equation of motion of an energy-driven AGN
wind outflow. The equation is derived based on three assumptions:
\begin{itemize}
\item Matter in the galaxy is distributed spherically symetrically;
\item The outflow is perfectly adiabatic;
\item The outflow does not have significant velocity gradients and
  therefore moves as a single entity.
\end{itemize}
The validity of these assumptions is discussed in Section
\ref{sec:caveats} below. The equation can then be derived using an
appropriate formulation of Newton's second law:
\begin{equation}
\frac{{\rm d}}{{\rm d}t}\left[M\dot{R}\right] = 4\pi R^2 P -
\frac{GM\left[M_{\rm b}+M/2\right]}{R^2}
\end{equation}
and the energy equation:
\begin{equation}
\frac{{\rm d}}{{\rm d}t}\left[\frac{3}{2}PV\right] =
\frac{\eta}{2}L_{\rm AGN} - P\frac{{\rm d}V}{{\rm d}t} - \frac{{\rm
    d}E_{\rm g}}{{\rm d}t},
\end{equation}
In these equations, $M(R)$ is the instantaneous swept--up gas mass
being driven out when the outflow contact discontinuity is at radius
$R$, $P$ is the expanding gas pressure and $M_{\rm b}$ is the mass of
the stars and dark matter within $R$ (these are left unmoved by the
outflow). The outflowing material is assumed to be distributed
approximately uniformly in a shell between $R$ and the outer shock at
$4R/3$ \citep[cf.][]{Zubovas2012ApJ}; this assumption has no effect on
the results presented here, as I am only considering integrated
quantities of the outflow. The energy equation expresses the change of
internal energy of the outflow (left-hand side) as a balance between
energy input by the AGN with luminosity $L_{\rm AGN}$, the $p$d$V$
work of outflow expansion and the work against gravity.

A detailed derivation of the equation of motion is presented in
\citet{Zubovas2016MNRASb}. Here, I just quote the final result:
\begin{equation} \label{eq:eom}
  \begin{split}
    \dddot{R} &= \frac{\eta L_{\rm AGN}}{M R} - \frac{2\dot{M}
      \ddot{R}}{M} - \frac{3\dot{M} \dot{R}^2}{M R} - \frac{3\dot{R}
      \ddot{R}}{R} - \frac{\ddot{M} \dot{R}}{M} \\ &
    +\frac{G}{R^2}\left[\dot{M} + \dot{M}_{\rm b} +
      \dot{M}\frac{M_{\rm b}}{M} - \frac{3}{2}\left(2M_{\rm
        b}+M\right)\frac{\dot{R}}{R}\right].
  \end{split}
\end{equation}
Here, $\dot{M}\equiv \dot{R}\partial M/\partial R$ and $\ddot{M}
\equiv \ddot{R}\partial M/\partial R + \dot{R} ({\rm d}/{\rm
  d}t)\left(\partial M/\partial R\right)$. In the rest of the paper, I
refer to the first term on the right hand side of equation
(\ref{eq:eom}) as the driving term, the next four terms as the kinetic
terms, and the terms involving the gravitational constant $G$ as the
gravity term.

This equation is then integrated numerically, using a specified
history of nuclear activity $L_{\rm AGN}\left(t\right)$, assuming a
central SMBH of mass $M_{\rm BH} = 2 \times 10^8 \;\msun$, giving an
Eddington luminosity $L_{\rm Edd} = 1.3\times 10^{46}$~erg/s. The
actual histories investigated are presented at the beginning of
Section \ref{sec:single}.

The matter distribution in the galaxy consists of two components. The
first component is an $M_{\rm h} = 6 \times 10^{11} \; \msun$ halo,
which has an NFW \citep{Navarro1997ApJ} density profile with virial
radius $r_{\rm vir} = 200$~kpc and concentration parameter $c = 10$
and is composed almost purely of dark matter and stars, with a gas
fraction of only $f_{\rm g,h} = 10^{-3}$. The second component is the
bulge, with a Hernquist density profile with scale radius $r_{\rm b} =
1$~kpc, total mass $M_{\rm b} = 4 \times 10^{10} \; \msun$ and a large
gas content $f_{\rm g,b} = 0.8$. The effect of this matter
distribution is that the outflow tends to accelerate significantly
when it moves outside $R \sim 3$~kpc, if the AGN is still active at
that time. The outflow starts at a radius $R_0 = 10$~pc with a
velocity $v_0 = 200$~km s$^{-1}$, which is approximately the expected
velocity dispersion in the bulge. The propagation of the outflow does
not depend strongly on the initial conditions \citep{King2011MNRAS}.

\subsection{Results - single activity episode} \label{sec:single}

\begin{figure*}
  \centering
    \includegraphics[trim = 0 0 0 0, clip, width=\textwidth]{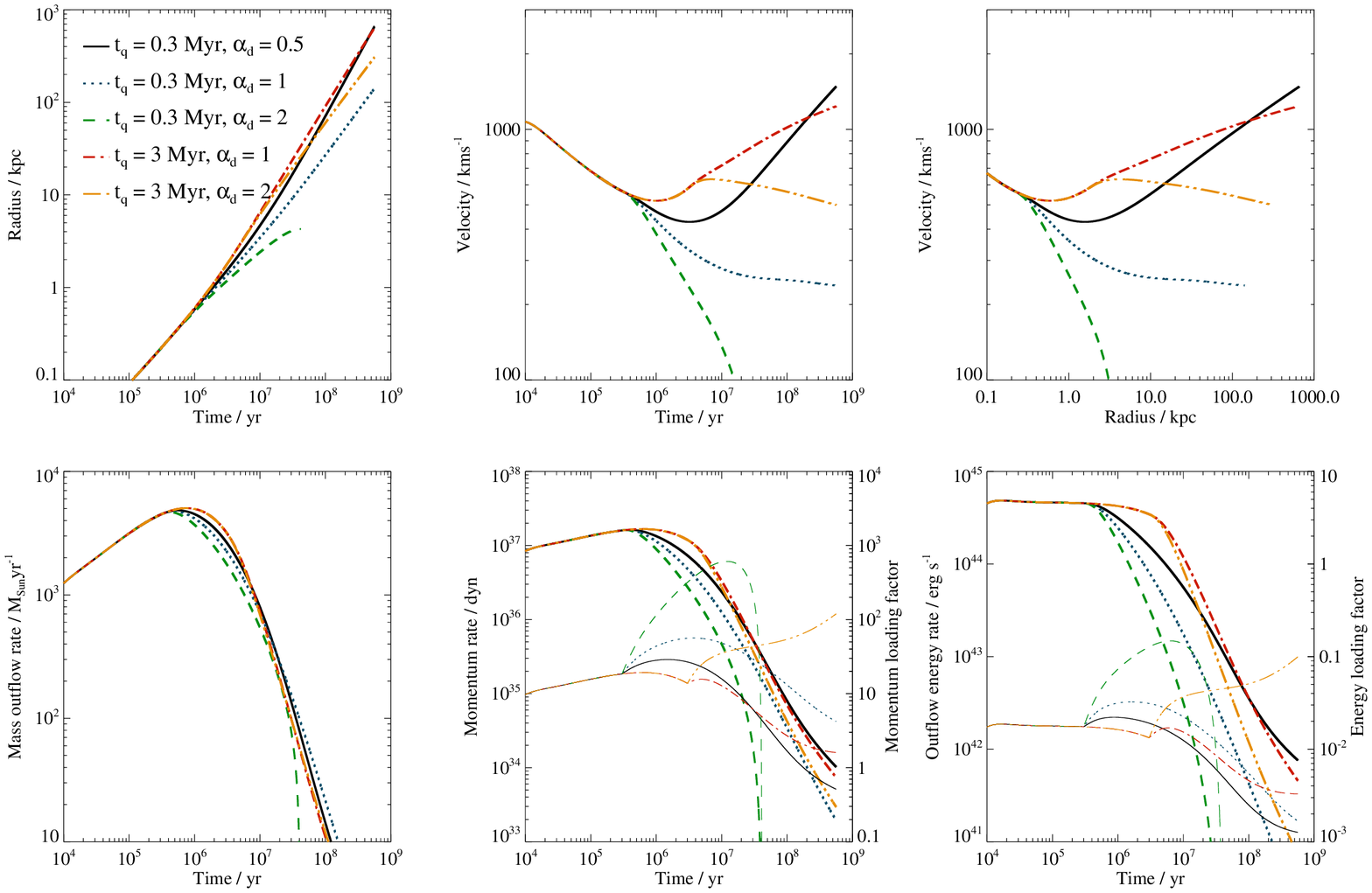}
  \caption{Same as Figure \ref{fig:expdrop}, but for simulations
    with power-law decay of AGN luminosity.}
  \label{fig:powerdrop}
\end{figure*}

\begin{figure*}
  \centering
    \includegraphics[trim = 0 0 0 0, clip, width=\textwidth]{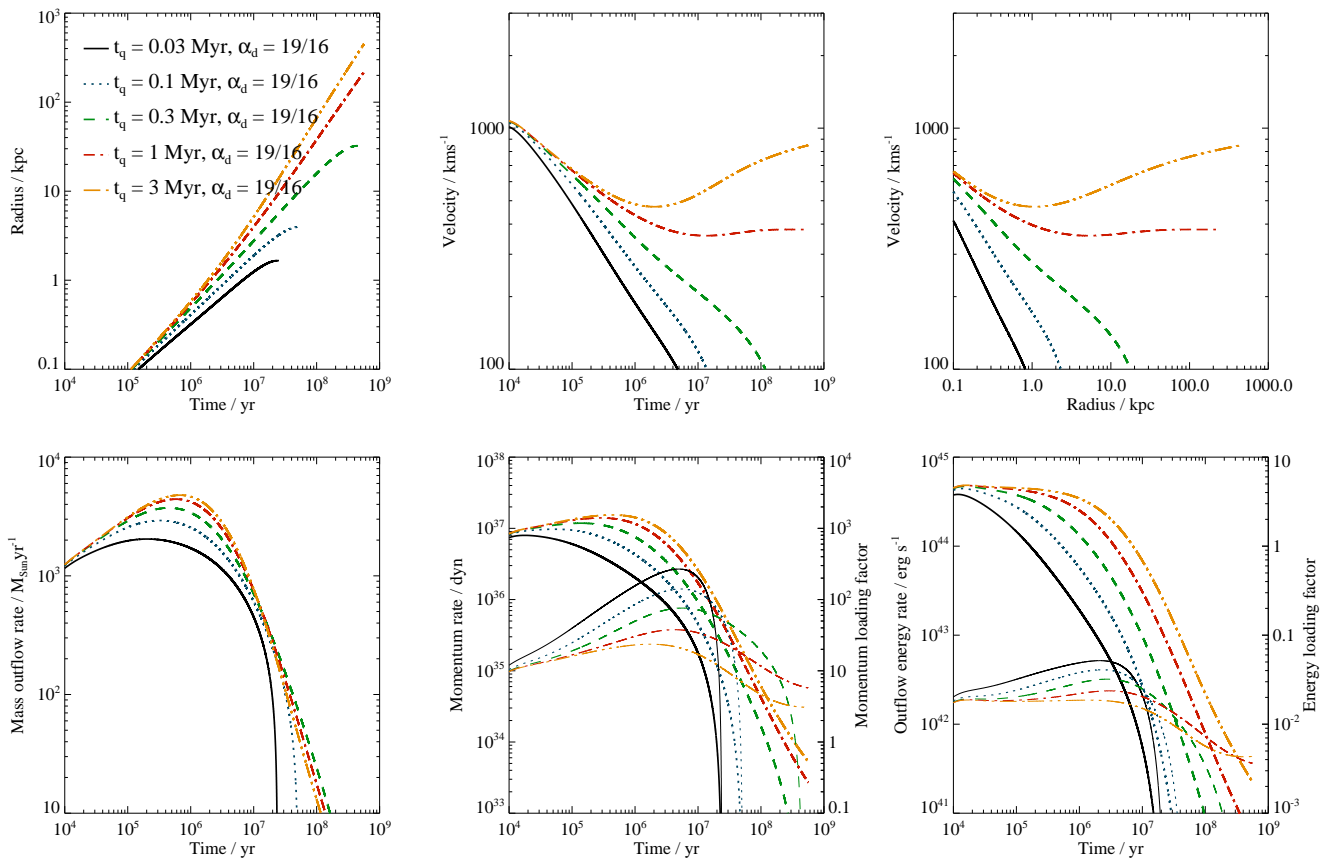}
  \caption{Same as Figure \ref{fig:expdrop}, but for simulations
    with extended power-law decay of AGN luminosity.}
  \label{fig:kingdrop}
\end{figure*}

\begin{figure*}
  \centering
    \includegraphics[trim = 0 0 0 0, clip, width=\textwidth]{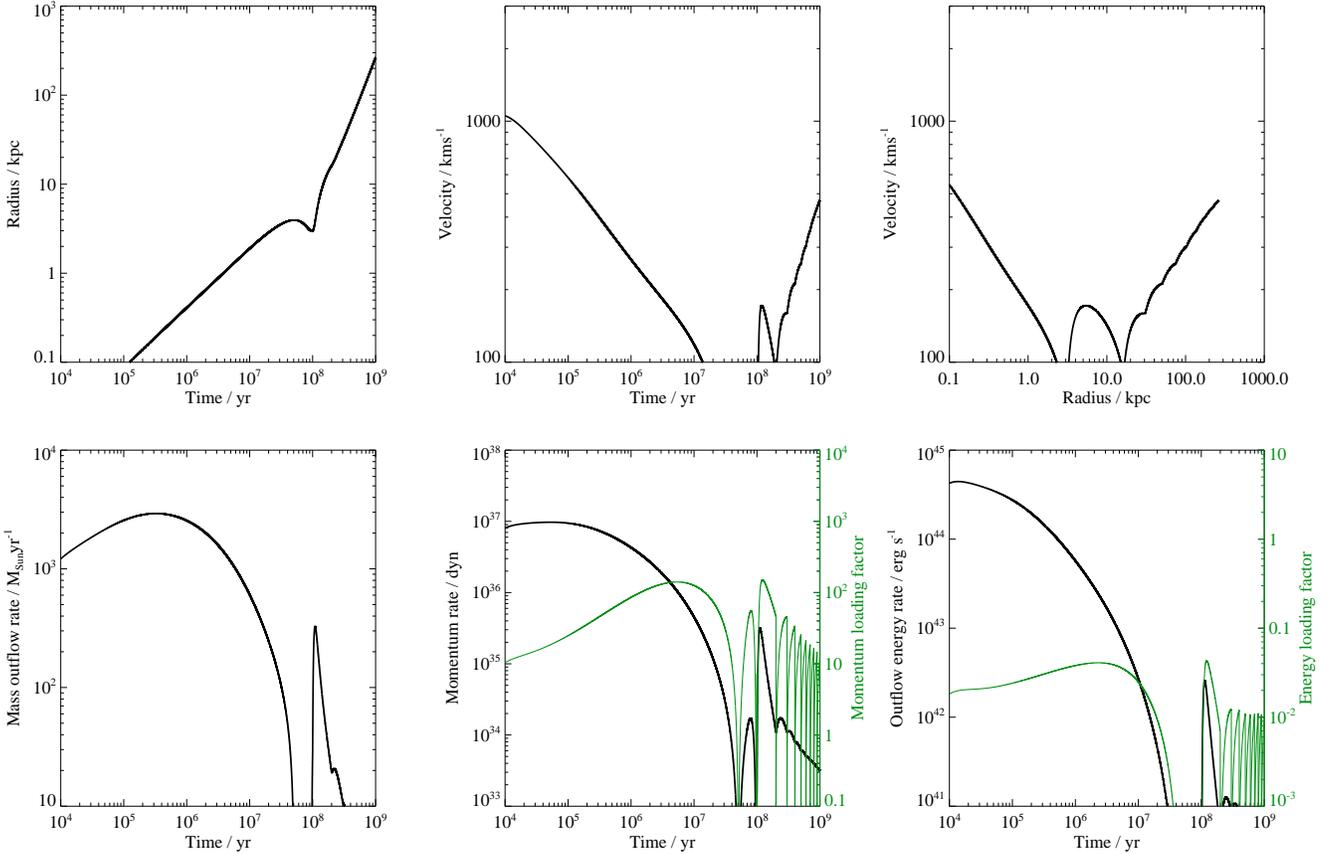}
  \caption{Outflow parameters in simulation with multiple AGN
    episodes, each of duration $t_{\rm q} = 10^5$~yr, separated by
    $t_{\rm rep} = 10^8$~yr, following an extended power-law decay of
    the AGN luminosity. Thin green lines in the last two panels show
    the momentum and energy loading factors.}
  \label{fig:king_multi}
\end{figure*}

\begin{figure}
  \centering
    \includegraphics[trim = 0 0 0 0, clip, width=0.45\textwidth]{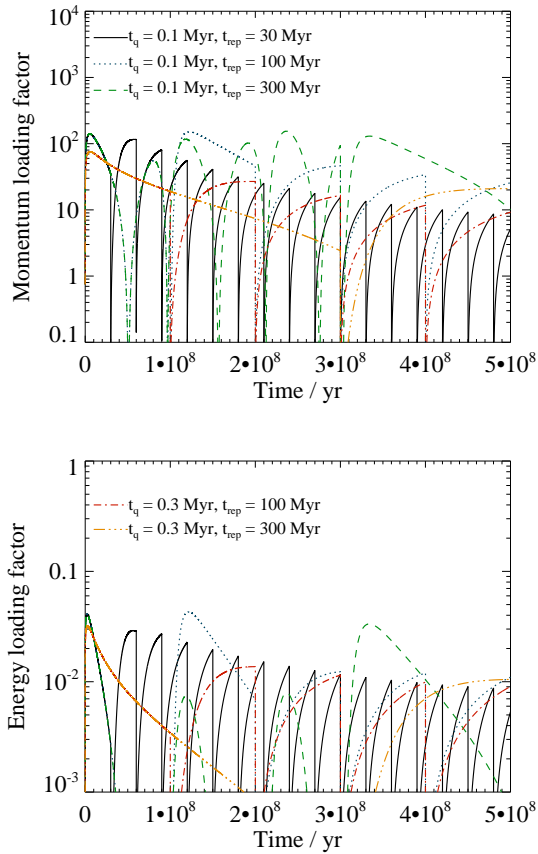}
  \caption{Momentum (top) and energy (bottom) loading factors in
    simulations with multiple AGN episodes, with different
    characteristic episode durations $t_{\rm q}$ and repetition
    timescales $t_{\rm rep}$, following an extended power-law decay of
    the AGN luminosity.}
  \label{fig:loadfac}
\end{figure}

\begin{figure}
  \centering
    \includegraphics[trim = 0 0 0 0, clip, width=0.45\textwidth]{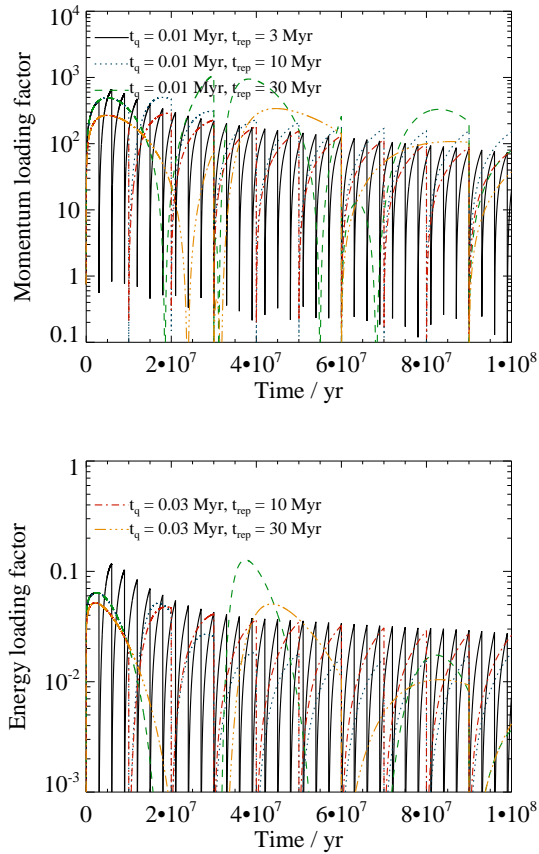}
  \caption{Same as Figure \ref{fig:loadfac}, but for AGN episodes with
    shorter characteristic durations.}
  \label{fig:loadfac_short}
\end{figure}

\begin{figure}
  \centering
    \includegraphics[trim = 0 0 0 0, clip, width=0.45\textwidth]{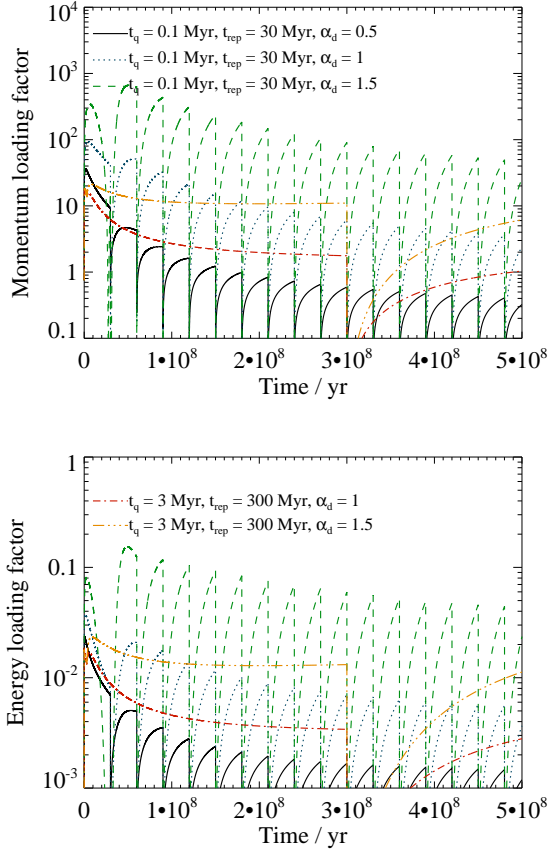}
  \caption{Same as Figure \ref{fig:loadfac}, but for AGN episodes with
    power law decay with different exponents $\alpha_{\rm d}$.}
  \label{fig:powerdrop_2}
\end{figure}

\begin{figure}
  \centering
    \includegraphics[trim = 0 0 0 0, clip, width=0.45\textwidth]{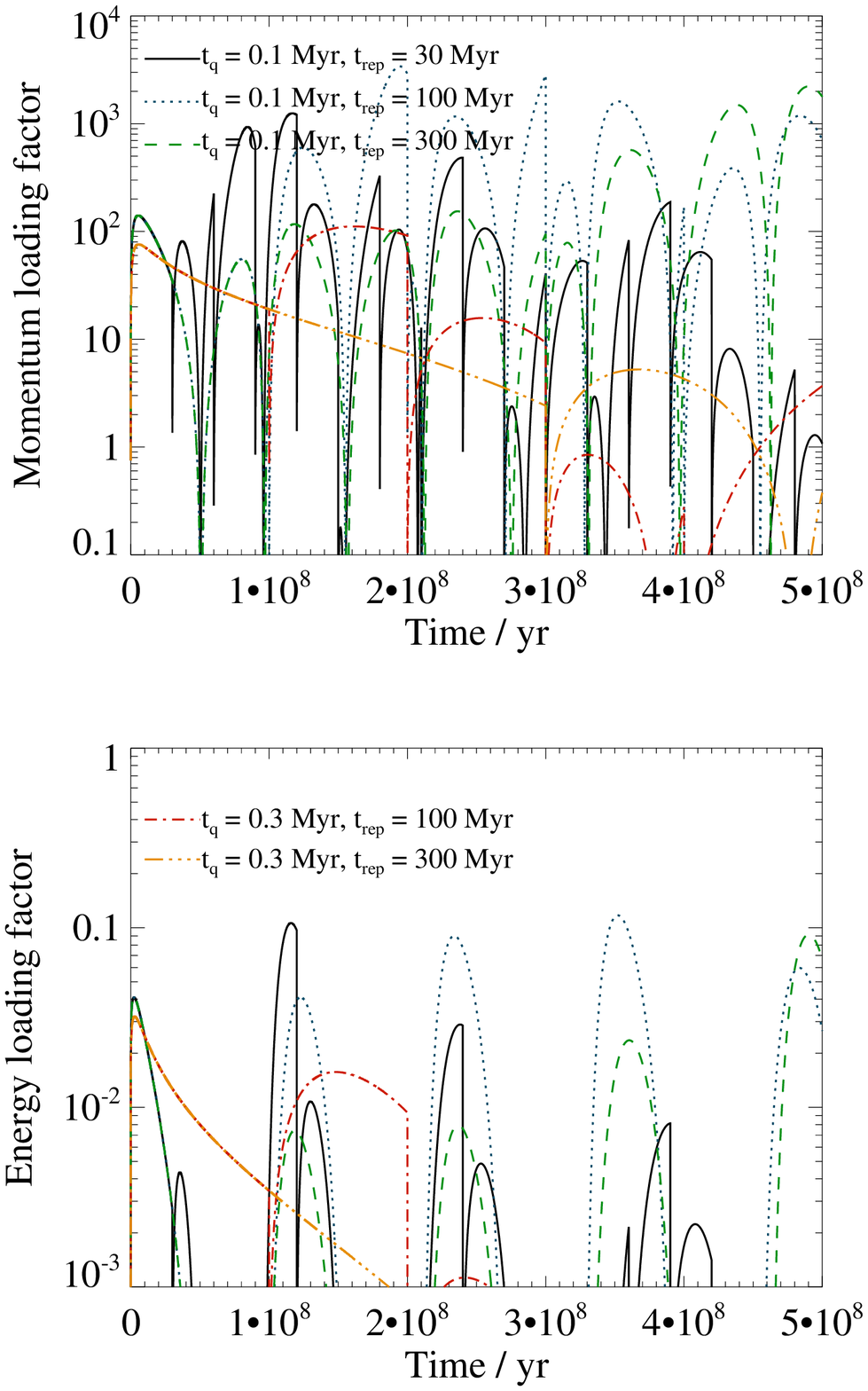}
  \caption{Same as Figure \ref{fig:loadfac}, but AGN episodes have
    progressively smaller initial luminosities, starting at $L_{\rm
      Edd}$ and decreasing by factor $3$ with each episode.}
  \label{fig:multiep_varlum}
\end{figure}

I present the results of numerical calculations of outflow
propagation, starting from the simplest possible AGN luminosity
history: a single event with luminosity $L = L_{\rm Edd}$, lasting for
a duration $t_{\rm q}$, after which the AGN turns off forever. This is
clearly an unrealistic situation, but it helps determine the
timescales over which the properties of a coasting outflow evolve.

Next, I consider three possibilities of AGN luminosity decay:
exponential, power-law and extended power-law. In the first case,
approximately following the model of \citet{Haehnelt1993MNRAS}, the
AGN luminosity is described as
\begin{equation}
  L_{\rm AGN}\left(t\right) = \begin{cases}
    L_{\rm Edd}, & t < t_{\rm q}, \\
    L_{\rm Edd}\; {\rm exp}\left(\frac{t_{\rm q}-t}{t_{\rm d}}\right), & t > t_{\rm q},
  \end{cases}
\end{equation}
where $t_{\rm d}$ is the luminosity decay timescale, a free parameter.

The power-law decay of AGN luminosity is described as follows:
\begin{equation}
  L_{\rm AGN}\left(t\right) = \begin{cases}
    L_{\rm Edd}, & t < t_{\rm q}, \\
    L_{\rm Edd}\left(\frac{t}{t_{\rm q}}\right)^{-\alpha_{\rm d}}, & t > t_{\rm q},
  \end{cases}
\end{equation}
where the exponent $\alpha_{\rm d}$ is a free parameter.

Finally, the extended power-law decay is based on the prediction by
\citet{King2007MNRAS}, which is given above in
eq. (\ref{eq:ext_powerlaw}). I repeat it here for ease of reference:
\begin{equation}
  L_{\rm AGN}\left(t\right) = L_{\rm Edd}\left(1+\frac{t}{t_{\rm q}}\right)^{-19/16}.
\end{equation}
With this prescription, the AGN remains in an active state, defined by
$L > 0.01 L_{\rm Edd}$ \citep[e.g.,][]{Shankar2013MNRAS}, for $t_{\rm
  ep} \simeq 47 t_{\rm q}$.  \citet{King2007MNRAS} use a
characteristic timescale $t_{\rm q} \sim $~a few times $10^5$~yr,
although a somewhat shorter timescale might be necessary to keep the
total duration of the AGN episode in line with observations
\citep{Schawinski2010ApJb, Ichikawa2016PASJ, Keel2017ApJ}.

All model AGN luminosity histories have a free parameter $t_{\rm q}$,
with exponential and power-law histories having one more free
parameter, respectively $t_{\rm d}$ or $\alpha_{\rm d}$. For
simulations with only one free parameter, I calculate outflow
propagation with $t_{\rm q} = 0.03, 0.1, 0.3, 1$ and $3$~Myr. For the
exponential decay simulations, I use $t_{\rm q} = 0.3$ and $3$~Myr and
$t_{\rm d} = 0.1, 0.3$ and $3$~Myr. For the power-law decay
simulations, I use $t_{\rm q} = 0.3$ and $3$~Myr and $\alpha_{\rm d} =
0.5, 1$ and $2$. As seen in the figures below, all outflows take $\sim
10^7$~yr or longer to reach $R \sim 10$~kpc, which is the largest size
of most observed outflows \citep{Spence2016MNRAS, Fiore2017A&A}, so
each of the model systems would undergo at least a few, and perhaps
several tens of, AGN episodes before the outflow would dissipate into
obscurity and be no longer detectable.

Figures \ref{fig:constdrop}, \ref{fig:expdrop}, \ref{fig:powerdrop}
and \ref{fig:kingdrop} show the evolution of main observable outflow
parameters with time in the four groups of models: immediate AGN
shutdown, exponential luminosity decay, power-law decay and extended
power-law decay, respectively. For the sake of clarity, I omit the
case $t_{\rm q} = 3$~Myr, $t_{\rm d} = 0.1$~Myr from the exponential
decay plots, because this case shows an even more extreme decay of
outflow parameters than $t_{\rm q} = 3$~Myr, $t_{\rm d} = 0.3$~Myr,
with a corresponding extreme increase in the loading
factors. Similarly, the case $t_{\rm q} = 3$~Myr, $\alpha_{\rm d} =
0.5$ is omitted from the power-law decay plots, because its evolution
is very similar to that of $t_{\rm q} = 0.3$~Myr, $\alpha_{\rm d} =
0.5$.

The parameters shown are outflow radius
$R$ (top left panels), outflow velocity $v$ (top middle panels),
velocity against radius (top right), mass outflow rate
$\dot{M}$(bottom left), momentum rate $\dot{M}v$ (bottom middle) and
kinetic energy rate $\dot{M}v^2/2$ (bottom right). In Figures
\ref{fig:expdrop}, \ref{fig:powerdrop} and \ref{fig:kingdrop}, the
last two panels also show the ratio between the outflow property in
question and the corresponding AGN property, $L_{\rm AGN}/c$ or
$L_{\rm AGN}$, showing the `loading factors' of the outflow.

In the immediate shutdown simulations (Figure \ref{fig:constdrop}),
the outflow velocity decays approximately as $t^{-\beta}$, with $\beta
\in \lbrace 0.3;0.6\rbrace$. The fastest decay occurs when the AGN
activity duration is shortest. This is caused by two effects: closer
to the centre of the galaxy, the potential well is deeper, requiring
more work to lift the gas, and the gas density is higher, stopping the
outflow more efficiently. The mass outflow rate decreases much faster
than velocity, and faster in the outflows inflated by longer AGN
activity episodes, because of the decrease in the gas density. Other
parameters are a function of $\dot{R}$ and $\dot{M}$. In particular,
the outflow momentum rate decay has a power-law exponent with
$\beta_{\rm \dot{p}} \in \lbrace 1.0;2.0\rbrace$, and the kinetic
energy rate has $\beta_{\rm \dot{E}} \in \lbrace 1.5;2.2\rbrace$. The
important observation here is that all these parameters decay
approximately as power-law functions. This suggests that a power-law
decay of AGN luminosity might preserve the AGN-outflow correlations,
while an exponential one might not.

Simulations with exponential AGN luminosity decay (Figure
\ref{fig:expdrop}) confirm this suspicion. Independently of the
precise value of the luminosity decay timescale $t_{\rm d}$, the
outflow momentum and energy rates decay far slower than the AGN
luminosity, leading to very large momentum and energy loading
factors. When these factors increase to extremely high values
($\dot{p}c/L_{\rm AGN} \ga 10^3$, $\dot{E}/L_{\rm AGN} > 1$), the
outflow might no longer be identified as AGN-driven. If there is a
powerful starburst happening in the galaxy \citep[perhaps even
  triggered by the same outflow, cf. ][]{Silk2005MNRAS,
  Zubovas2013MNRASb}, the outflow might be classified as driven by
star formation, especially at $t > 10^7$~yr, when the momentum and
energy rates have decreased significantly and are similar to the
typical starburst-driven outflow parameter values
\citep{Cicone2014A&A, Geach2014Natur, Heckman2015ApJ}. Given that the
outflow material cools down and some of it might not be detected, the
estimated rates may be smaller than the true ones and the outflow
properties might begin to resemble those of star-formation driven
outflows earlier than $10^7$~yr after the beginning of the AGN
phase. At intermediate times, however, outflows should be observed and
correctly classified as AGN-driven; for example, the model with
$t_{\rm q} = 0.3$~Myr and $t_{\rm d} = 0.1$~Myr (black solid line)
spends $\sim 0.9$~Myr at $L > 0.01 L_{\rm Edd}$, out of which
$0.4$~Myr, i.e. almost half the time, is spent with $\dot{p}c/L_{\rm
  AGN} > 100$. This fraction is even higher if $t_{\rm d}/t_{\rm q}$
is higher. The lack of observed outflows with very large momentum or
energy loading factors indicates that exponential decay of AGN
luminosity is rare, if not non-existent.

Simulations with power-law decay of AGN luminosity (Figure
\ref{fig:powerdrop}) show momentum and energy loading factors that
generally fall within the range of observed values. Momentum loading
in most cases stays between $10$ and $10^2$ for more than $10^7$~yr
after the AGN starts to dim. Only when the AGN activity is short and
decay very rapid (green dashed lines) does the momentum loading factor
increase to more than $10^2$, but this happens only at a time when the
AGN luminosity has decreased by a factor $10^2 - 10^3$, so it is
questionable whether such an outflow would be identified as AGN-driven
in the first place. Another effect worth noting is that a slow decay
of AGN luminosity can sometimes lead to the outflow breaking out of
the galaxy anyway (black solid and red dot-dashed lines), since it
provides enough energy to the gas for it to escape from the
gravitational potential well.

Finally, the extended power-law simulations (Figure
\ref{fig:kingdrop}) result in even less variation of the momentum and
energy loading factors. The kinetic energy rate of the outflow is in
all cases $<10\%$ of the AGN luminosity, while the momentum rate only
becomes higher than $100 L_{\rm AGN}/c$ in the two simulations with
the shortest activity duration $t_{\rm q}$. I conclude that this type
of AGN luminosity evolution function is the best, of the three tested
ones, at explaining the continued existence of tight correlations
between AGN luminosity and outflow parameters even during the AGN
decay phase.

\subsection{Results - repeating activity episodes}

Using the extended power-law model of AGN luminosity decay, I test the
effect of the AGN experiencing several episodes of activity. At first,
for simplicity, I assume that each episode reaches the same maximum
luminosity equal to the Eddington value. The full luminosity function
is then characterised by two free parameters: $t_{\rm q}$ and the time
between the beginnings of two successive activity episodes, $t_{\rm
  rep}$.

With this luminosity function, the duty cycle $f_{\rm d}$ of the AGN,
if defined as the fraction of time for which the AGN is brighter than
some fraction $f$ of its Eddington luminosity, is given by
\begin{equation} \label{eq:fd}
  f_{\rm d} = \left(f^{-16/19}-1\right)\frac{t_{\rm q}}{t_{\rm rep}} =
  10^{-3}\left(f^{-16/19}-1\right)\frac{t_{\rm q,5}}{t_{\rm rep,8}},
\end{equation}
where in the second equality I scale $t_{\rm q}$ and $t_{\rm rep}$ to
$10^5$ and $10^8$~yr, respectively. Using a canonical AGN definition
of $f = 0.01$ \citep[e.g.,][]{Shankar2013MNRAS} gives $f_{\rm d} =
0.047$, in reasonable agreement with observations
\citep{Kauffmann2003MNRAS, Xue2010ApJ}.

Figure \ref{fig:king_multi} shows plots of the major outflow
parameters in the simulation with $t_{\rm q} = 10^5$~yr and $t_{\rm
  rep} = 10^8$~yr. The outflow stalls and begins collapsing back
before the second AGN episode; this adiabatic oscillation was removed
from single-episode plots, because it is unlikely to occur in reality
due to fragmentation of the outflowing gas. Subsequent AGN episodes
push the outflowing gas out of the galaxy. Crucially, the momentum and
energy loading factors (green lines in the last two panels) generally
stay within the observationally-constrained bounds: $\dot{p}c/L_{\rm
  AGN} < 100$, $\dot{E}/L_{\rm AGN} < 0.05$. When the outflow moves
beyond a few tens of kpc, the mass outflow rate becomes negligible and
the outflow is unlikely to be detected. This happens after only 2-3
AGN episodes, supporting the notion that AGN outflows clear gas out of
galaxies in a comparatively short time \citep{Zubovas2012ApJ}.

Variations of the momentum and energy loading factors are shown in
Figures \ref{fig:loadfac} and \ref{fig:loadfac_short} for outflows
with different values of $t_{\rm q}$ and $t_{\rm rep}$. I use a linear
scale for time, cut off at $t_{\rm max} = 500$~Myr in Figure
\ref{fig:loadfac} and at $t_{\rm max} = 100$~Myr in Figure
\ref{fig:loadfac_short}, because subsequent episodes generally show
very similar results. In the five simulations with $t_{\rm q} \geq
0.1$~Myr (Figure \ref{fig:loadfac}), the momentum loading factors do
not exceed $200$, generally staying below a few tens. The energy
loading factors are always below $5\%$ and do not rise above $2\%$ in
most AGN episodes. In the simulations with shorter $t_{\rm q} <
0.1$~Myr (Figure \ref{fig:loadfac_short}), both momentum and energy
loading factors rise to higher values, reaching as much as $10^3$ for
momentum loading. However, these outflows have very low velocities,
$v_{\rm out} < 200$~km/s, and therefore are unlikely to be
distinguished from the random motions of gas in the host galaxy.

The extended power-law simulations all have the same exponent
governing AGN luminosity decay, so I ran several tests with simple
power-law AGN luminosity evolution in order to determine the
importance of the exponent. The results are qualitatively similar to
those of the extended power-law simulations and single-episode
power-law simulations (Figure \ref{fig:powerdrop}). Some example
results are given in Figure \ref{fig:powerdrop_2}. The exponent of the
AGN luminosity decrease is a key factor governing the maximum
momentum/energy loading factors reached. With exponent $\alpha_{\rm d}
= 0.5$, the AGN luminosity decreases so slowly that the momentum
loading factor stays below $\sim 5$ after $t = 30$~Myr, and the energy
loading factor is $<1\%$. With $\alpha_{\rm d} = 1.5$, the momentum
loading factor increases to more than $100$, while the energy loading
factor is $\sim 10\%$. When $\alpha_{\rm d} = 1$, both the momentum
and energy loading factors are very similar to those in the extended
power law simulation. This should be expected, because the extended
power law simulation has AGN luminosity decreasing with a power-law
exponent $-19/16$, i.e. close to $-1$, at late times.  If the AGN
activity episodes are very long, both momentum and energy loading
factors remain small, despite a rapid decrease in AGN luminosity.

\subsection{Results - varied-luminosity episodes}

Finally, I run a simulation with multiple AGN episodes of different
starting luminosity. I take the characteristic durations $t_{\rm q}$
and repetition timescales $t_{\rm rep}$ to be the same as in the first
group of simulations with multiple Eddington-limited extended power
law episodes (presented in Figure \ref{fig:loadfac}). The only
difference between the episodes is that only the first activity
episode has a starting luminosity $L = L_{\rm Edd}$. Each subsequent
episode has a starting luminosity three times lower than the previous
one, but not lower than $0.03 L_{\rm Edd}$. This represents a
situation where the first activity episode drives an outflow which
removes most of the gas from the central parts of the galaxy, and
subsequent activity episodes are unable to fuel the AGN very
efficiently. Each subsequent episode removes some more of the leftover
gas, and so the maximum AGN luminosity decreases further.

The resulting momentum and energy loading factors are presented in
Figure \ref{fig:multiep_varlum}. In simulations with $t_{\rm q} =
0.1$~Myr, the momentum loading factors rise to rather large values,
$>10^2$ and even $>10^3$ in some instances. This suggests that in some
cases, old outflows illuminated by a low-luminosity AGN can have very
large formal momentum-loading factors, although this happens because
the outflow momentum rates are inherited from previous, more luminous,
AGN episodes. The low-luminosity AGN is unable to accelerate the
outflow significantly, and so the gas dynamics is similar to that of a
stalling outflow, characterised by very low velocities, $v < 50$~km/s
at these late times (an example of an outflow with low velocity and
high momentum loading can be seen in Figure \ref{fig:king_multi}, at
$10^8$~yr$ < t < 2\times 10^8$~yr). Such outflows would be very
difficult to detect and would appear as a radial anisotropy of gas
motions in the host galaxy, rather than as a
kinematically-distinguishable outflow. If detected, they might fill
the extreme end of the momentum loading - velocity diagram, such as
the right panel of Figure 2 in \citet{Fiore2017A&A}.

Due to the low velocity of these outflows at late times, they tend to
have very low energy loading factors, even when the AGN itself has
very low power. This is consistent with observations, which show a
super-linear relation between AGN luminosity and outflow kinetic
energy, i.e. low-luminosity AGN tend to host under-energetic outflows
\citep[right panel of Figure 1 in ][]{Fiore2017A&A}.

\section{Discussion} \label{sec:discuss} 

\subsection{Correlation observability}

In the simulations with extended power-law AGN luminosity evolution,
the momentum and energy loading factors approximately follow the
observed correlations ($\dot{p}c/L_{\rm AGN} \sim 20$, $\dot{E}/L_{\rm
  AGN} \sim 0.05$) during the first AGN episode. During later
episodes, outflow energy and momentum rates are lower than during the
first one, and fall below the observed correlations. This finding
suggests that outflows with momentum and energy rates lower than
expected from current correlations might be common. These outflows
would be difficult to observe, explaining why the properties of
currently known outflows follow the correlations.

Another implication of the findings is that low-luminosity AGN should
show a higher variety of outflow properties than high-luminosity
ones. This happens because high-luminosity AGN and quasars are more
likely to be driving the present outflow, while low-luminosity ones
might be only illuminating an outflow inflated by an earlier, perhaps
more powerful, episode. A low-luminosity AGN might also be in the
fading phase of a more powerful outflow-driving episode, resulting in
outflow momentum and energy being higher than expected given the
current AGN luminosity.

Many outflows that are observed in galaxies without an AGN might in
fact be `fossil' AGN outflows, left over from an earlier AGN episode
\citep{King2011MNRAS}. These outflows are commonly assumed to be
driven by star formation, and often show good correlations with SFR,
as if they were driven by the momentum produced by newborn stars
\citep[e.g.,][]{Cicone2014A&A}. Given that the AGN episode might
enhance the star formation rate on timescales of $10^7-10^8$~yr
\citep{Zubovas2013MNRASb, Zubovas2016MNRASb}, it would be interesting
to determine whether any correlation might be expected between the
properties of the AGN outflow and the triggered star formation rate in
the galaxy disc. This investigation is, however, beyond the scope of
the present paper.

\subsection{Outflows as constraints on long-term AGN variability}

Both observational \citep[][]{Schawinski2015MNRAS} and theoretical
\citep[][]{King2015MNRAS} constraints on the duration of AGN activity
suggest that the typical AGN episode duration is $t_{\rm q} \sim
10^5$~yr. The definitions used in these two papers are, however,
somewhat different. The observational constraint of
\citet{Schawinski2015MNRAS} encompasses a full AGN episode, from the
rapid `turn-on' to complete shutdown (meaning at least that $L < 0.01
L_{\rm Edd}$). The theoretical argument of \citet{King2015MNRAS}
provides a timescale of accretion episode evolution, which leads to
the full AGN episode being $\sim 47$ times longer (see eq. \ref{eq:fd}
above). Therefore, the observational constraint suggests shorter AGN
episodes than theoretical arguments do.

Timescales over which AGN fade also differ when estimated
observationally and theoretically. Observations suggest very rapid AGN
fading, $t_{\rm d} \sim 10^4$~yr \citep{Schawinski2010ApJb,
  Keel2017ApJ}, but the viscous evolution of an accretion disc predict
a power-law dropoff at late times \citep{King2007MNRAS}, and the AGN
has a rather low luminosity for most of the duration of its episode.

The properties of AGN outflows can help provide some additional
constraints and potentially explain the difference between the two
estimates. Comparing the results of multi-episode simulations (Figures
\ref{fig:loadfac} and \ref{fig:loadfac_short}), I find that outflows
inflated by more rapidly flickering AGN are slower, and have higher
typical momentum loading than those inflated by more slowly flickering
AGN. However, if the outflow has already progressed far from the
nucleus, i.e. there is little gas in the central parts of the galaxy,
then both rapidly and slowly flickering AGN can continue to push it
further without producing unrealistic momentum loading factors. This
suggests that AGN and their outflows progress differently depending on
the gas fraction of the host galaxy:
\begin{itemize}
\item In galaxies with large gas fractions in the spheroid, typically
  at high z, AGN are fed by large gas reservoirs, which are resilient
  to feedback, and thus the typical AGN episode duration is long,
  $t_{\rm ep} \sim 47 t_{\rm q} \sim 5$~Myr. Such AGN are capable of
  inflating large and massive outflows. It is important to note that
  \citet{Schawinski2015MNRAS} did not include such galaxies in their
  analysis, since their samples were limited to $z < 0.4$.
\item In galaxies with low gas fraction, typically at low z, AGN are
  fed by intermittent gas reservoirs, and thus the typical AGN
  episode duration is short, $t_{\rm ep} \sim t_{\rm q} \sim
  0.1$~Myr. Such rapidly-flickering AGN can only inflate large
  outflows because there is little gas in the host galaxy.
\end{itemize}

This scenario is, of course, highly idealised, and numerical
hydrodynamical simulations would be required in order to test whether
a difference in host galaxy gas fraction can produce such a
qualitative difference in outflow properties. The point remains,
however, that gas-rich galaxies should be able to feed their central
SMBHs for long and almost continuous episodes, while gas-poor galaxies
are likely to have more intermittent AGN episodes. This distinction
can help explain why local AGN do not generally appear to be driving
outflows \citep{Nedelchev2017arXiv}. The fact that local AGN flicker
on short timescales suggests that they are fed rather
intermittently. This can mean that the whole host galaxy is mostly
devoid of gas, and any outflow inflated by the AGN is difficult to
detect due to faintness. Alternatively, the galaxy might have a lot of
gas, but the immediate surroundings of the AGN can be depleted, and
AGN outflows are very slow and therefore indistinguishable from the
random gas motions in the host galaxy.

Based on the hypothesis presented above, I predict that if galaxies
are divided into two populations - those with observed outflows and
those without - galaxies with observed outflows will show noticeably
longer typical AGN episode durations than galaxies without
outflows. This prediction can be tested as more data on outflows is
collected. Another prediction is that the typical AGN episode duration
should be higher in high-redshift galaxies than in the local Universe,
but this is almost a corollary of the previous one.

\subsection{Caveats and improvements} \label{sec:caveats}

The model presented above, and the conclusions stemming from its
results, depend upon several assumptions, mainly about the driving
mechanism of the outflow. They have been stated in Section
\ref{sec:nummodel}.

The assumption of perfect spherical symmetry results in a singular,
well defined, outflow radius. This is obviously an idealised
situation, whereas in reality, the ISM of the host galaxy is highly
non-uniform, and the outflow only propagates in some directions, while
potentially stalling and collapsing in others. This clearly makes any
interpretation of observations more difficult. On the other hand,
given that the properties of observed outflows agree very well with
theoretical predictions based on such a simplified model, I am
confident that variations in outflow properties due to time
variability of AGN can be distinguished.

The assumption that the outflow is perfectly adiabatic is also not
precisely correct. Although the shocked wind is almost certainly
adiabatic, the shocked ISM generally cools rapidly
\citep{Zubovas2014MNRASa, Richings2017arXiv}. This results in a
density stratification in the outflow, with cold clouds coalescing
from the ISM and being subsequently pushed only by the momentum of the
wind. The energy, momentum and pressure retained in the hot diffuse
gas then diminishes. This might help maintain the observed
correlations as the AGN fades. In addition, and during subsequent AGN
episodes, this fragmentation might contribute to the observed
gas-to-radiation pressure ratios, which are much lower than the simple
adiabatic outflow model would predict \citep{Stern2016ApJ}.  The lower
energy and momentum rates of real outflows would make it more
difficult to use abnormal correlations, or lack thereof, to constrain
possible AGN activity histories. Nevertheless, this constrainment
should be possible if the motion of all gas phases is accounted
for. However, accounting for radiation-pressure effects in the driving
of outflows \citep[e.g.,][]{Ishibashi2012MNRAS} might be necessary in
order to get the full picture.

Finally, the model assumes that the wind mass outflow rate is
identical to the SMBH accretion rate. This would generally not be the
case. In gas-rich systems, the AGN may be fed at a super-Eddington
rate, leading to much denser winds than predicted here, while in
gas-poor systems, winds might also be diffuse. A denser wind would
move with a lower velocity, assuming that it maintains an electron
scattering optical depth $\tau_{\rm es} = 1$, i.e. remains purely
momentum-driven. It would also have a lower energy rate, which would
lead to it inflating a less powerful outflow. Such an outflow might
remain within the observational bounds of correlations for a longer
period while the AGN fades. Conversely, diffuse outflows would move
with higher velocities and energy rates, inflate more powerful
outflows, which would more strongly constrain the possible
evolutionary history of AGN luminosity. Investigating these issues in
detail is beyond the scope of the present paper, but I hope to model
them with a hydrodynamical numerical model of varying AGN outflows.

\section{Conclusion} \label{sec:concl}

In this paper, I presented 1D numerical simulations of AGN outflow
propagation following different AGN luminosity histories. The goal of
this work was to determine which AGN luminosity histories preserve the
observed AGN-outflow correlations during the fading of the AGN and
during subsequent AGN episodes, which do not drive the outflow from
the start. The main results are the following:
\begin{itemize}
\item If the AGN fades exponentially, there is a non-negligible period
  of time for which the outflow would be seen as having abnormally
  large momentum and energy loading factors.
\item If the AGN fades as a power-law with exponent $\alpha_{\rm d}
  \sim 1$, the correlations between AGN luminosity and outflow
  properties are preserved. Such an evolutionary history is a
  natural consequence of viscous disc evolution.
\item Subsequent AGN episodes following the first one only increase
  the outflow momenum and energy to values smaller than predicted by
  the correlations, even if the maximum AGN luminosity of these
  episodes is smaller than that of the first one.
\item As a result, I predict that weaker AGN outflows will be detected
  in the future, with momentum and energy rates lower than given by
  current correlations; the currently-known correlations are only an
  upper limit to the full variety of outflow properties.
\item Outflow properties can be used to constrain AGN episode
  properties: for example, I predict that local galaxies that produce
  massive outflows have on average longer-lasting individual AGN
  episodes than galaxies without outflows.
\end{itemize}

The possibility of using AGN outflows to constrain the properties of
past activity episodes makes them an important tool for understanding
galaxy evolution. This is one more piece of evidence showing the
outflows can serve as dynamical footprints allowing us to investigate
the past behaviour of galaxies.

\section*{Acknowledgments}

This research was funded by grant no. LAT-09/2016 from the Research
Council Lithuania.


\end{document}